# Human-Aware Power Restoration for Fair Outage Experience in Distribution Systems

Majid Dehghani, Arastoo H salimi, Student *Member, IEEE*, Hamidreza Nazaripouya, *Senior Member, IEEE*

***Abstract*—This paper proposes a novel methodology for human-aware and fair service restoration in power distribution networks, explicitly accounting for the customer experience of outage duration. The complexity of this problem stems from the inherently unpredictable and stochastic nature of power outage events. Traditional approaches often oversimplify the problem by treating failures as deterministic, overlooking the lived experiences of customers and the true uncertainty of outage patterns. In contrast, the proposed method incorporates the probability of potential failures to guide a customer-aware and fairness-driven resource allocation, ensuring that restoration is not only fast but also perceived as fair from the customer's perspective. To achieve this, a spatially distributed, adaptive, and scalable partitioning policy is designed to balance restoration time across all failure locations, promoting consistency and equity in the outage experience. Next, an adaptive and distributed repair crew dispatch algorithm is proposed to accelerate service restoration while ensuring that no customer segment is disproportionately affected. The framework leverages a Receding Horizon (RH) optimization algorithm to dynamically minimize total restoration time amid randomly occurring outages in both space and time. Simulation results on modified 69-bus distribution networks under stochastic outage conditions demonstrate the model's effectiveness in delivering socially fair and customer-sensitive restoration outcomes.**

## NOMENCLATURE

*Sets and Indices*

| | |
|---|---|
| $m/n$ | Indexes for damaged components and storage depots |
| $\beta_s$ | Set of buses which are substations |
| $i/j$ | Indices for partitions |
| $l$ | Indices for buses |
| $dp$ | Designated return point for the repair teams |
| $k$ | Index for the distribution line linking two buses |
| $t$ | Index for time |
| $K(.,l)$ | Set of lines with bus $l$ as the to bus |
| $K(l,.)$ | Set of lines with bus $l$ as the from bus |

*Parameters*

| | |
|---|---|
| $d(dp,m)$ | The distance from the depot to the damaged component. $m$ |
| $nc$ | Number of crews at the depot |
| $np$ | Number of partitions |
| $P(t)$ | The repair crew positions at time $t$ |
| $p^D_{l,t}/q^D_{l,t}$ | Active and reactive load at bus $l$ during time $t$ |
| $tr_{m,n}$ | Duration of travel for a crew from $m$ to $n$ |
| $\alpha$ | Weight factor |
| $W$ | The weight of a power diagram |
| $g$ | The positions of a power diagram |
| $f(x)$ | The density function over $A$ |
| $FD$ | Failures Density |
| $r_m$ | Expected time needed by a crew to repair damaged component $m$ |

*Variables*

| | |
|---|---|
| $AT_m$ | Time of arrival for a crew at the damaged component $m$ |
| $B_{i,j,t}$ | Binary variable equals 1 if $i$ is the parent bus of $j$ and 0 otherwise. |
| $p^{DG}_{l,t}/q^{DG}_{l,t}$ | Generated active and reactive power by the DG at bus $l$ |
| $p^L_{k,t}/Q^L_{k,t}$ | Active and reactive power flow on line $k$ at time $t$ |
| $u^L_{k,t}$ | Binary variables for the operational status of the line $k$ at time $t$ |
| $RF_{m,t}$ | Binary variable indicating the time damaged component $m$ is repaired |
| $x_{m,n}$ | Binary variable showing if crew moves from damaged component $m$ to $n$ |
| $y_m$ | Binary variable indicating whether a crew visited damaged component $m$ |
| $z_{m,t}$ | Binary variable for the availability of damaged component $m$ at time $t$ |
| $WT_j$ | The wait time for failure $j$ |
| $\rho_{l,t}$ | Status of the load connection at bus $l$ |

## I. INTRODUCTION

DISTRIBUTION systems play a critical role in delivering electricity from utilities to consumers, but they are susceptible to disruptions caused by equipment failures. As extreme weather events become more frequent, power interruptions are expected to rise, increasingly disrupting daily life and essential services [1]. Therefore, it is essential for utility companies to promptly identify affected areas, effectively manage repair crews, and restore electrical services [2][3]. With limited resources available, the central challenge is not only to allocate them efficiently, but to do so in a way that reflects a human-aware approach—one that prioritizes fairness in the customer outage experience across all communities. A critical yet often overlooked aspect of restoration planning is the need to ensure restoration fairness from customer's perspective, recognizing that prolonged outages can profoundly affect human quality of life. While conventional approaches focus on minimizing total outage durations or maximize load restoration, they can unintentionally lead to disproportionate wait times for certain customers, especially those in geographically dispersed or underserved communities. To avoid leaving any community significantly disadvantaged, especially during large-scale and evolving outage events, restoration strategies must move beyond efficiency alone and incorporate fairness as a core design principle. This shift is essential for navigating the complexity of real-world outages, where the dynamic and uncertain nature of failures—and their unpredictable emergence—can significantly affect the human experience if not addressed with a human-aware approach.

There have been considerable efforts in developing power system restoration techniques [4][5][6]. Some researchers emphasize the importance of determining the shortest route to the damaged components for effective emergency repairs. Consequently, they have utilized the vehicle routing problem (VRP) framework for identifying optimal routes. Reference [7] has addressed power system emergency repairs as a VRP while considers multiple vehicles. This study introduces a computational framework for determining repair vehicle routes, utilized by a power company in Santiago, Chile. However, the determination of repair task priority, vehicle location, and repair duration relied heavily on the dispatcher's experience. Reference [8] has studied the power restoration vehicle routing problem by decoupling the power-restoration and vehicle-routing optimizations. However, optimizing solely for the shortest path does not necessarily maximize the load restored within minimal time after an outage. To this end, depending on the situation, higher priority may be given to repairing equipment that restores larger loads. Reference [9] used a mixed-integer linear programming (MILP) technique to combine VRP with the power system emergency repair criteria in order to reduce lost load while expediting the restoration of loads.

Additionally, several researchers have attempted to distribute the workload among repair crews by geographically clustering damaged equipment to reduce the computational complexity of the restoration problem [10], [11], [12]. Reference [10] has proposed agglomerative clustering-based network partitioning strategy to identify appropriate partitions for parallel repair. Reference [11] presents a spectral clustering methodology that leverages the network's physical and inherent properties to restore isolated sections of the network. The method constructs an undirected, edge-weighted graph based on distances and incorporates constraints through edge-weight modifications. A clustering-based technique is presented in [12] to minimize power outages in distribution systems by utilizing microgrids and increasing the likelihood of restoring critical loads after natural disasters. The method identifies restoration routes and reduces outages during normal operations by strategically leveraging distributed energy resources (DERs) in accordance with the distribution network topology. Microgrids are formed based on load density, and optimal locations for DER deployment are identified to enhance restoration effectiveness.

While service restoration for distribution systems has been widely studied, most approaches treat damages as deterministic [13], [14], assuming that the location and timing of failures are known and static. In practice, however, damage occurs unpredictably, and new repair requests can emerge at any time during the restoration process, especially during extreme weather events. This unpredictability complicates the restoration problem, especially when striving for fairness. Ensuring fair restoration for all customers—so that no customer waits disproportionately longer than another—is a critical challenge, which has not been adequately studied. While partitioning methods such as clustering have been proposed, they fall short when future damages are uncertain, making it difficult to fairly allocate resources and maintain restoration fairness from a customer perspective. It is essential that restoration strategies uphold human-aware fairness, ensuring that all individuals—regardless of location—receive timely and dignified access to electricity during critical times.

To address this limitation, this paper models the restoration problem within a comprehensive framework that incorporates human-aware fairness into both the partitioning of service regions and the routing of repair crews. By accounting for the stochastic nature of failure occurrences—including their uncertain location, timing, and repair duration—the proposed strategy dynamically adapts to evolving conditions. The objective is to minimize the expected system time (i.e., the sum of waiting and service durations), while promoting restoration fairness from a customer perspective, so that all users receive comparable service regardless of when or where the outage occurs.

This paper proposes an adaptive, distributed, and scalable partitioning policy designed to ensure human-aware fairness in service delivery, providing consistent restoration experiences to all customers, regardless of their location. The Receding Horizon (RH) policy is then employed to determine the optimal route for addressing the fair restoration problem under stochastic failure conditions within each partition.

In summary, the paper makes the following contributions:

- Developing a distinctive restoration framework that models the stochastic nature of failure occurrences, effectively addressing the uncertainty inherent in such events, a factor often neglected in previous approaches.
- Designing an innovative, adaptive, and spatially distributed policy to equitably partition the restoration workspace through a human-centered framework, addressing restoration fairness from the customer's perspective, a challenge often overlooked in traditional approaches.
- Designing an adaptive and distributed algorithms for coordination of multiple repair crews, optimizing their routes and interactions to improve overall system performance.

The power system restoration formulation adopted in this work builds upon well-established and validated models reported in the literature. Rather than redefining the underlying network constraints, this paper focuses on developing a human-aware and fairness-driven restoration framework that operates on top of these models. In this sense, the proposed approach is designed to be model-agnostic, allowing it to be integrated with different power system formulations while maintaining its effectiveness under stochastic outage conditions.

The remainder of this paper is organized as follows: the background and problem description are given in Section II. Section III presents a human-aware restoration framework that enables fair partitioning to enhance fairness in customer service restoration. Section IV presents the Receding Horizon (RH) policy, which provides a solution to the fair restoration problem under stochastic failure conditions. The simulation results are presented in Section V, and conclusion remarks are provided in Section VI.

## II. Preliminaries

### A. Problem Description

In restoration problems, multiple repair crews are typically employed to repair damaged components. One approach is to divide the affected area into smaller partitions, with each crew assigned to a specific partition for more effective repair. The challenge stems from the non-deterministic nature of failures, which complicates the equivalent distribution of workload among crews. Unlike deterministic scenarios, where the fixed and known number of damaged components can be equally shared among crews, the unpredictable location and timing of failures require incorporating the stochasticity of future failures into the partitioning process. Fig. 1 illustrates an example of both existing and potential future failures. If future failures align with the red crosses, the red partitioning strategy can ensure fair repair wait times for customers. However, if the density of future failures follows the blue crosses, the red partitioning becomes unfair. As shown in the figure, partition 1 experiences considerably more failures compared to partitions 5 or 6. To address this inequality, a different partitioning needs to be implemented, for instance by merging partitions 5 and 6, and further dividing partition 1. This underscores the importance of incorporating the stochastic nature of future failures into the partitioning process to ensure human-aware fairness in restoration times.

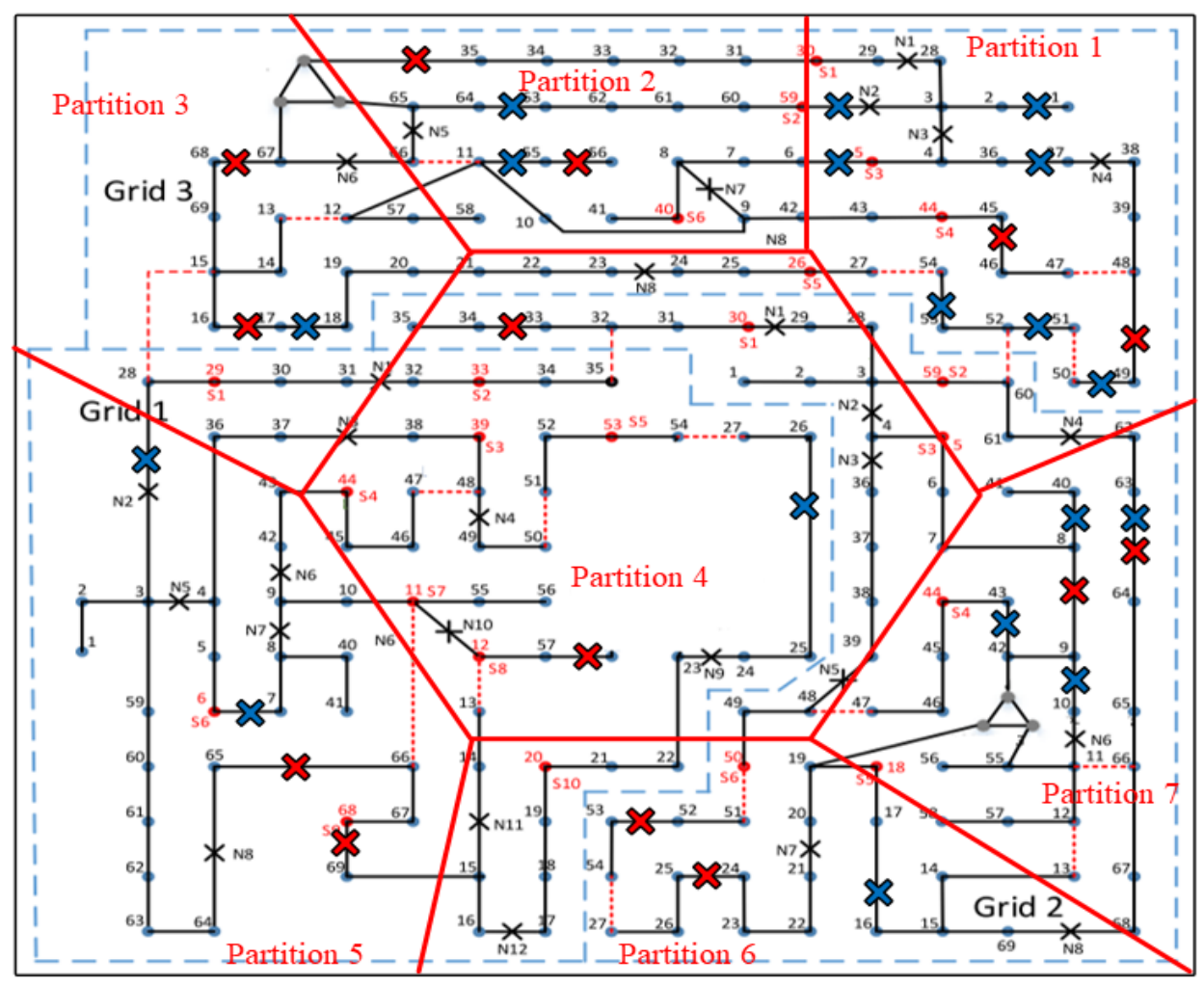

**Fig. 1.** Existing and potential future failures

In this paper, a mathematical model for fair restoration is proposed, where failures occur stochastically within a service territory. The model includes the following: *i*) a group of repair crews are assigned to repair damaged components; *ii*) a stochastic modeling of failures in both time and space to capture the uncertainty of outage events; and *iii*) an objective to minimize the expected total repair time—including both waiting and service times—while ensuring fair restoration for all affected customers over an infinite horizon.

The repair crew positions at time $t$ can be represented by $P(t) = \{p_1(t), \dots, p_{nc}(t)\}$. These crews can move freely within area $A$ at a constant velocity $v$. Failures occur according to a homogeneous (time-invariant) spatio-temporal Poisson process with time intensity $\lambda \in R_{>0}$[15]. The location of failures $\{X_j;\ j \geq 1\}$ are independently and identically distributed (i.i.d.) over $A$ according to a spatial density function $f : A \rightarrow {R^2}_{>0}$. For clarity, since $f : A \rightarrow {R^2}_{>0}$, $f$ is a vector-valued function such that $f(x) = [f_1(x), f_2(x)]$, where $x = [x_1, x_2]$ represents the coordinates in $A$. The location of a failure becomes known at the time of occurrence; hence, crews know the locations of all failures that occurred before time $t$, while the locations of future failures form an i.i.d. sequence. The density $f$ is satisfied according to (1):

$$P[X_j \in A_i] = \int_{A_i} f(x)\, dx \quad \forall A_i \subset A \text{ and} \int_A f(x)\, dx = 1 \quad (1)$$

The number of initial failures is assumed to be a random variable. At each failure location $X_j$, crews spend a specific amount of time $s_j$ to provide on-site service. A failure is repaired and removed from the system after a crew's on-site work is finished. The failure load factor is defined as $FL = \lambda \times s/nc$.

The time between the occurrence of failure $j$ and the completion of service by the crew is referred to as the system time ($T_j$) for failure $j$, where failures numbered in ascending order based on their occurrence times. The formula $WT_j = T_j - s_j$ represents the wait time for failure $j$. The steady-state system time is represented by $\overline{T} = lim\ sup_{j\rightarrow\infty} E[T_j]$. If the expected number of failures within the system is always uniformly bounded, then the crew routing policy is said to be stable. It is assumed throughout this study that $FL < 1$, which is an essential condition for the existence of a stable policy.

This research suggests a two-step method to tackle the issue of fair restoration and repair in distribution systems. Initially, a spatially distributed and adaptive policy is designed to create fair partitions by considering the existing failures and the probability of future failures, where the probability of future failures is derived from statistical analysis of historical utility data on failure locations. A partitioning policy is an algorithm that divide a bounded workspace $A \subset R$ into $nc$ completely disjoint areas $A_i$ for $i \in I_{nc}$, $I_{nc} = \{1,2,\dots,nc\}$, depending on the number of available crews and their locations. Each crew in the restoration problem is assigned to a subregion, denoted as $A_i$, and is responsible for addressing all failures within that subregion. Crew $i$'s workload $f_{A_i}$ for a subregion $A_i \subset A$ is modelled as $f_{A_i} = \int_{A_i} f(x)\, dx$, where $f(x)$ is spatial density function across $A$. Considering this structure, load-balancing seeks to distribute workload evenly among the $nc$ subregions. In other words, it aims to calculate fair partitions over $A$ so that, for every $i$, $f_{A_i} = f_A/nc$.

Partitioning helps break down the routing repair crews (RRC) problem into several smaller dynamic vehicle routing problems (DVRPs) from a single large DVRP. This decomposition significantly reduces the complexity of computational optimization problems.

A distributed and adaptive unbiased policy is used in the next step to find the optimal route for the fair restoration problem. This strategy does not explicitly require knowledge of global network topology; instead, it relies only on local information sharing between nearby crews. The goal is to reduce the expected restoration time for failures including the total of the wait and service times.

## III. Distribution System Repair and Restoration Operation Framework

This section presents a comprehensive framework for the restoration problem, with a focus on restoration time and fairness. Damage assessment leverages consumer reports, prediction algorithms, failure detection methods, and aerial surveys. After that, the utility uses this data to assign repair teams. These crews must handle failures that may occur stochastically in both time and location during progressive extreme weather events. Each crew has a designated route to travel from failures to their depots after finishing all their assigned tasks. The goal is to maximize the restored loads while minimizing the total expected repair time equitably for all customers over an infinite time horizon.

This section provides a detailed description of the fair partitioning strategy and the mathematical formulation of the load restoration problem.

### A. Fair Partitioning

To design a restoration strategy that distributes workloads fairly, it is necessary to formalize the concept of “fair partitions” in the service territory. Simple distance-based partitions (e.g., nearest depot) do not account for uneven outage densities and may result in long delays for some customers. To capture fairness mathematically, geometric partitioning tools from computational geometry—Voronoi and Power diagrams—are employed to model service regions based on both crew positions and failure density. The following definitions and theorems provide the mathematical foundation needed to describe fair partitions and to guarantee their existence. In particular, the definitions formalize these concepts, while the theorems and lemmas guarantee that the proposed gradient-based method converges to fair partitions. .

***Definition 1 (fair Partitions)***: Given a measurable function $f: A \rightarrow R^2{}_{>0}$, an $nc$-partition $\{A_i\}_{i=1}^{nc}$ is considered fair with respect to $f$ if $\int_{A_i} f(x)\, dx = \frac{1}{nc}\int_A f(x)\, dx$ for every $i \in I_{nc}$.

***Definition 2***: Voronoi diagram $V(G) = \left(V_1(G), \cdots, V_{nc}(G)\right)$ of $A$ is defined as (2) where $G$ is an ordered collection of distinctive points $(g_1, \cdots, g_{nc})$.

$$V_i(G) = \{x \in A \mid \|x - g_i\| \leq \|x - g_j\|, \forall j \neq i, j \in I_{nc}\} \quad (2)$$

$V_i(G)$ is called the Voronoi cell or region of dominance for the $i$-th power point in this context, and $G$ is called the set of power points of $V(G)$. The Voronoi diagram $V(G)$ of a set $A$ partitions the space into regions $V_i(G)$ as shown in Fig.2, where each region $V_i$ contains all points in $A$ such as $F_m$ that are closer to $g_i$ than to any other point in $G$.

The issue with using a Voronoi diagram arises when the density of failures in one cell is significantly higher than in others, resulting in an unfair partitioning. In such cases, as shown in Fig. 3, partition $i$ has 8 failures, while some neighboring partitions have none. This uneven distribution results to longer wait times in partition $i$ compared to others, leading to unfair partitions. To address the inequity, the power diagram—an extended version of the Voronoi diagram—is introduced. This allows for more flexible partitioning that leads to a more fair distribution of workloads.

***Definition 3***: A Power diagram extends the concept of Voronoi diagrams by associating each partition with a specific coefficient, referred to as a "weight". This weight determines its dominance over a subregion. Let $W = (w_1, \cdots, w_{nc})$ be the distinctive weight assigned to each point $g_i \in G$, where $i \in I_{nc}$. Then, the power distance is described as follows:

$$d_P(x, g_i; w_i) = ||x - g_i||^2 - w_i \quad (3)$$

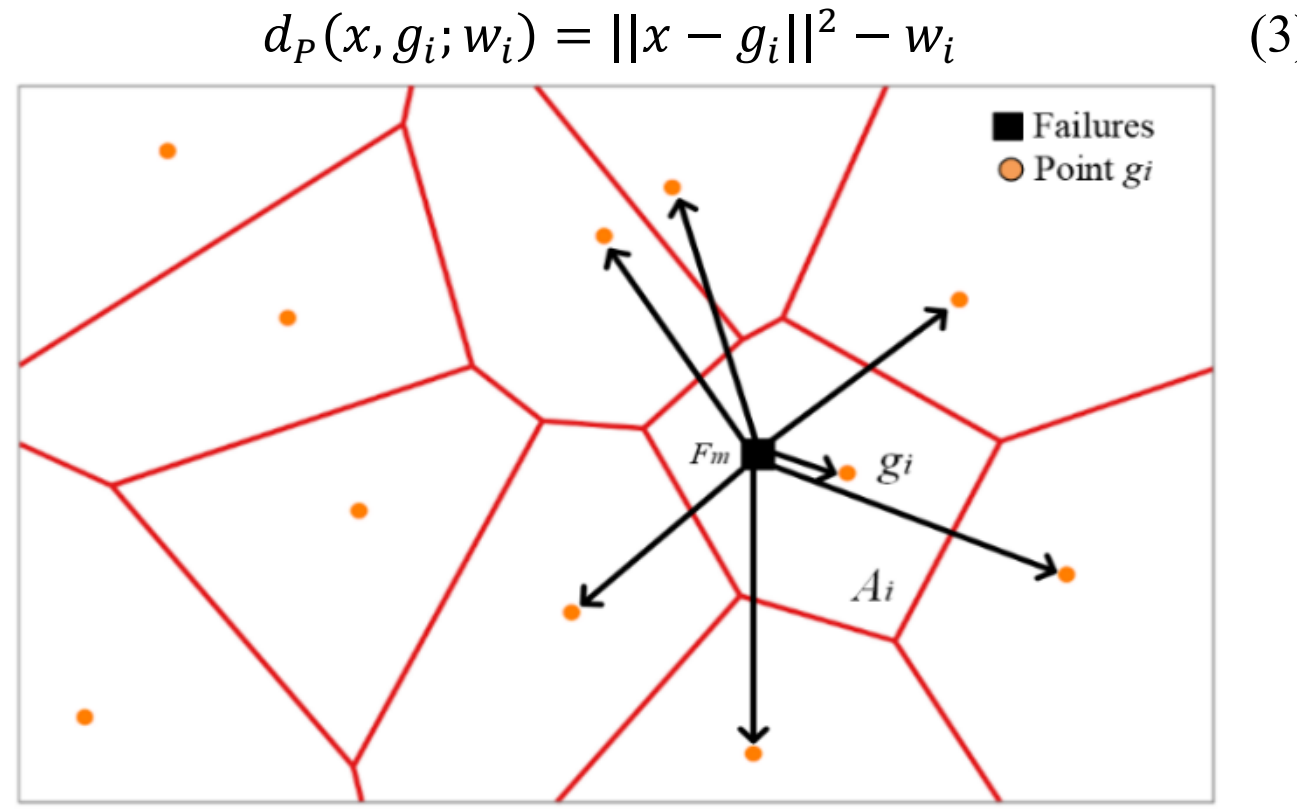

**Fig. 2.** Partitioning according to the Voronoi diagram

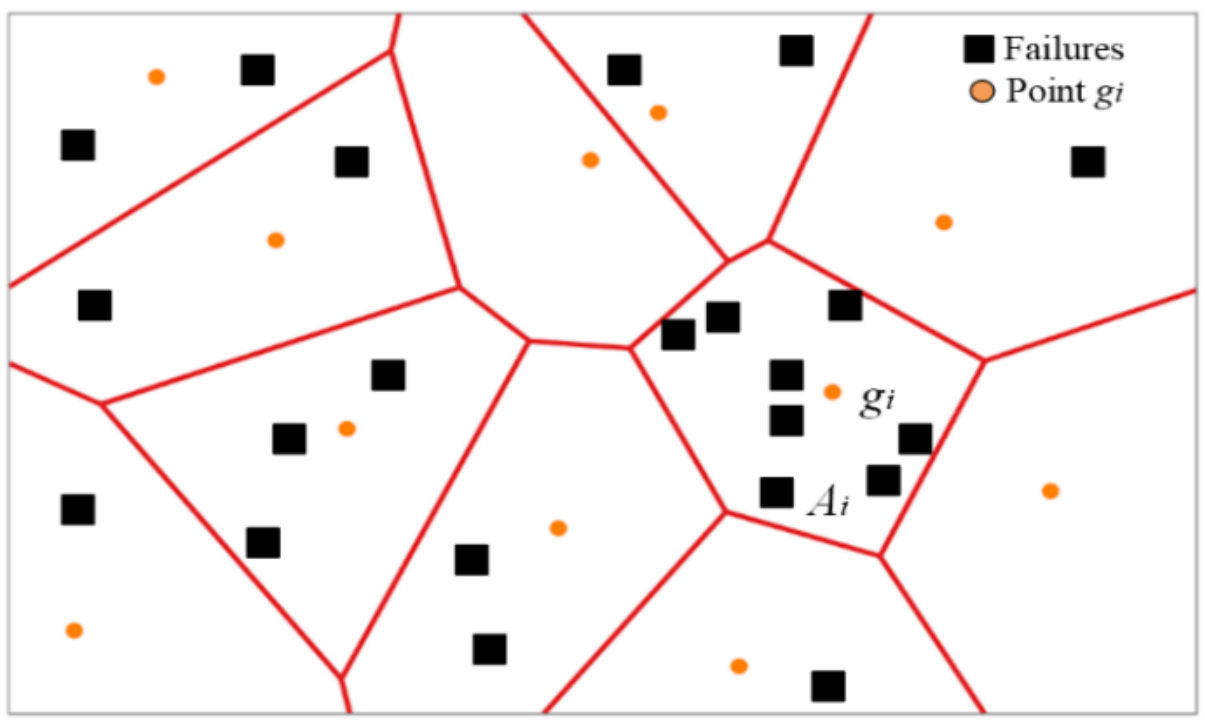

**Fig. 3.** Partitioning according to the Voronoi diagram with failures

The pair $(g_i, w_i)$ is referred to as a power point. The set of power points is denoted as $G_W = \{(g_1, w_1), \cdots, (g_{nc}, w_{nc})\}$. Like the Voronoi diagram, the Power diagram $V(G_W) = \left(V_1(G_W), \cdots, V_{nc}(G_W)\right)$ of $A$, produced by the power points $((g_1, w_1), \cdots, (g_{nc}, w_{nc}))$, is declared as follows [16]:

$$V_i(G_w) = \{x \in A \mid \|x - g_i\|^2 - w_i \leq \|x - g_j\|^2 - w_j, \forall j \neq i, j \in I_{nc}\} \quad (4)$$

Each region $V_i(G_w)$ is called the power cell or region of dominance of the $i$-th power point. The positions and weights of the power point $(g_i, w_i)$ are indicated by the terms $g_i$ and $w_i$, respectively. Remarkably, the power diagram and the Voronoi diagram of $A$ coincide when all weights are equal. For each $i \neq j$, the bisector of $(g_i, w_i)$ and $(g_j, w_j)$ is defined as:

$$b\left((g_i, w_i), (g_j, w_j)\right) = \{x \in A \mid (g_j - g_i)^T x = \frac{1}{2}(\|g_j\|^2 - \|g_i\|^2 + w_i - w_j)\} \quad (5)$$

The line segment $\overline{g_i g_j}$ is bisected by the bisector

$b\left((g_i, w_i), (g_j, w_j)\right)$ through the point $g_{ij}^*$, which is given by

$$g_{ij}^* = \frac{\|g_j\|^2 - \|g_i\|^2 + w_i - w_j}{2\|g_j - g_i\|^2}(g_j - g_i) \tag{6}$$

This final property is essential for the remainder of the paper. It demonstrates that the bisector between two power points can be freely adjusted by appropriately modifying the weights to achieve fair partitions. The following theorem ensures that an fair Power diagram exists for any set of distinct points $G = (g_1, g_{nc})$ with respect to any arbitrary measure $f$.

***Theorem 1***: Let density function $f$ be a measure across $A$. Let $G = (g_1, \cdots, g_{nc})$ denote the locations of distinct points within $A$ where $1 \le nc < \infty$. A Power diagram that is fair with respect to $f$ is produced by the power points $(g_1, w_1), \dots, (g_{nc}, w_{nc})$ given weights $w_i,\ i \in I_{nc}$. Additionally, if there is a vector of weights $w^*$ that results in an fair partition, then the set of all weight vectors that cause a fair partition is $W_t^* = \{w^* + t[1, \dots, 1]\}$, where $t \in R$ [17]. Interested readers can read [17] for proof. ■

According to the ***Theorem 1***, there is always a Power diagram that is fair with respect to $f$. Therefore, in this paper, an adaptive and distributed policy is designed to enable a team of crews to achieve an equipartition configuration by adjusting the weights $w_i,\ i \in I_{nc}$ .

Let $(g_i, w_i) \in A \times R$ represent the power point where $g_i = p_i$, indicating that the power point's position corresponds with the crew's position. The power cell $V_i = V_i(G_W)$ is then used to determine the partition for crew $i$, where $G_W = \{(g_1, w_1), \dots, (g_{nc}, w_{nc})\}$. The key idea is to enable the weights of the power points ($w$) to follow a spatially-distributed gradient descent law (while maintaining the positions of the power points fixed) such that a fair partition is reached.

Assume that each power point's position is distinct, that is, $g_i \ne g_j$ for $i \ne j$. Describe the set:

$$S = \{(w_1, \dots, w_{nc}) \in R \mid f_{A_i} > 0\ \forall\, i \in I_{nc}\} \tag{7}$$

In other words, the set encompasses all weight vectors so that every dominant region has a non-zero measure.

The locational optimization function is introduced:

$$H_A(W) = \sum_{i=1}^{nc}\left(\int_{A_i(W)} f(x)\, dx\right)^{-1} = \sum_{i=1}^{nc} f_{A_i(W)}^{-1} \tag{8}$$

where $W = (w_1, \dots, w_m)$, and $f_{A_i(W)}$ is workload for region $A_i$. The following lemma demonstrates that there exists a global minimum for $H_A$ with a specific set of weights $W = (w_1, \dots, w_m)$ that are the weights for achieving fair partitioning.

***Lemma 1***: A global minimum of $H_A$ is a vector of weights that produces a fair power diagram [16].

***Proof:*** Consider the relaxation of the minimization problem:

$$\min_{f_1,\dots,f_{nc}} \sum_{i=1}^{nc} f_i^{-1} \tag{9}$$

$$\text{s.t}\quad \textstyle\sum_{i=1}^{nc} f_i = a > 0,\ f_i > 0, i \in I_{nc}$$

where equality constraint arises from the fact that $\sum_{i=1}^{nc} \int_{A_i(W)} f(x)\, dx = \int_A f(x)\, dx = a$, where $a$ is a positive constant, given that $f$ is a density function defined over the domain $A$. Using Lagrange multipliers, it can be demonstrated that the global minimum for this relaxation occurs when $\int_{A_i(W)} f(x)\, dx = \frac{1}{nc}\int_A f(x)\, dx$ for all $i$. This ensures that the partitions are fair in accordance with ***Definition 1***. Since ***Theorem 1*** confirms the existence of a vector of weights in $S$ that creates a fair power diagram, it can be concluded that this vector is a global minimum of $H_A$. ■

Up to this point, ***Theorem 1*** and ***Lemma 1*** have demonstrated the existence of a set of weights $W = (w_1, \dots, w_m)$ that represent the global minimum of $H_A$ and achieve fair partitioning. Next, the focus is on calculating these weights.

A spatially distributed algorithm for fair partitioning is proposed. Assume that $\dot{w}_i = u_i$ describes the first-order dynamical behavior of the weights of the power points. Adopt the weight $w_i$ to follow the gradient descent provided by (10) and treat $H_A$ as an objective function to be minimized. To be more specific, the control law is as follows:

$$u_i = -\frac{\partial H_A}{\partial w_i}(W) \tag{10}$$

The partition $A(W) = \{A_1, \dots, A_{nc}\}$ is continuously updated until they achieve fairness.

**Discussion:** The gradient descent update law in (10) is employed because of its distributed and scalable nature, which is essential in large-scale distribution networks. Unlike centralized optimization approaches that require global information and can be computationally prohibitive, the gradient descent formulation enables each repair center (represented by a power point) to update its assigned region based only on local information exchanges with neighboring repair centers. This localized structure significantly reduces communication and computational burdens, while still guaranteeing convergence to a fair partition as shown by Lemma 1. Moreover, gradient-based methods are well-suited for dynamic environments where outage patterns evolve over time; the continuous update law allows the partitions to adapt in real-time. Alternative approaches, such as solving a global MILP for partitioning, would not scale to large systems or stochastic, time-varying failures, making gradient descent particularly attractive in this context.

***Theorem 2***: Assume that for any $i \ne j$, the power point locations are distinct, that is, $g_i \ne g_j$. Given the spatial density function $f$, the function $H_A$ is continuously differentiable on $S$, where for each $i \in I_{nc}$:

$$\frac{\partial H_A}{\partial w_i} = \sum_{j \in N_i} \frac{1}{2\gamma_{ij}}\left(\frac{1}{f_{A_j}^2} - \frac{1}{f_{A_i}^2}\right)\int_{\Delta_{ij}} f(x)\, dx \tag{11}$$

where $\gamma_{ij} = \| g_j - g_i \|$. Furthermore, the critical points of $H_A$ are weight vectors that yield a fair Power Diagram.

To prove ***Theorem 2***, it is required to transform the domain of the integral form a two-dimensional area into a one-dimensional line. Equation (12) is the result of classic divergence theorems [18]. Assume that the region $\Omega = \Omega(y) \subset A$ has a well-defined border $\partial\Omega(y)$ for all $y$, and depends smoothly on $y \in R$. If a density function over $A$ is denoted by $f$, then

$$\frac{d}{dy}\int_{\Omega(y)} f(x)\, dx = \int_{\partial\Omega(y)}\left(\frac{dx}{dy}\cdot n(x)\right) f(x)\, dx \tag{12}$$

where the unit outward normal to $\partial\Omega(y)$ is $n(x)$, the derivative of boundary points with respect to $y$ is indicated by $dx/dy$, and the scalar product of vectors $v$ and $w$ is indicated by $v.w$.

***Proof of Theorem 2:*** Given power point positions are distinct (i.e. $g_i \neq g_j$ for $i \neq$ j), this guarantees the existence of non-degenerate bisectors and prevents undefined or overlapping regions, which are necessary for constructing a valid partition of the domain. Since the motion of a weight $w_i$ only impacts power cell $A_i$ and its adjacent cells $A_j$ for any $j \in N_i$, where $N_i$ is a set of all neighbors of partition $i$, then:

$$\frac{\partial H_A}{\partial w_i} = -\frac{1}{f_{A_i}^2}\frac{\partial f_{A_i}}{\partial w_i} - \sum_{j\in N_i}\frac{1}{f_{A_j}^2}\frac{\partial f_{A_j}}{\partial w_i} \tag{13}$$

Equation (12) provides the means to determine how the value of an integral changes in response to a variation in the domain of integration. The boundary of $A_i$ can be expressed as $\partial A_i = \cup_j \Delta_{ij} \cup B_i$. Here, $\Delta_{ij} = \Delta_{ji}$ represents the boundary between $A_i$ and $A_j$. Additionally, $B_i$ denotes the boundary between $A_i$ and $A$ (if it exists, otherwise $B_i = \emptyset$). Consequently, the $\frac{\partial f_{A_i}}{\partial w_i}$ in (13) is calculated as follows:

$$\frac{\partial f_{A_i}}{\partial w_i} = \sum_{j\in N_i}\int_{\Delta_{ij}}\left(\frac{\partial x}{\partial w_i}\cdot n_{ij}(x)\right)f(x)\ dx + \underbrace{\int_{B_i}\left(\frac{\partial x}{\partial w_i}\cdot n_{ij}(x)\right)f(x)\ dx}_{0} \tag{14}$$

where $n_{ij}$ is defined as the unit outward normal to $\Delta_{ij}$ , and thus, $n_{ij}$=-$n_{ji}$. The second term is trivially equal to zero if $B_i = \emptyset$. It is also equal to zero if $B_i \neq \emptyset$; since $\frac{\partial x}{\partial w_i}\cdot n_{ij}(x) = 0$. For each point $x \in \Delta_{ij}$, (5) is differentiated with respect to $w_i$. This allows us to obtain the scalar product between the derivative of boundary points and the unit outward normal to the border in (14). To this end, one can have:

$$\frac{\partial x}{\partial w_i}\cdot\left(g_j - g_i\right) = \frac{1}{2} \tag{15}$$

Substituting (15) into (14):

$$\frac{\partial f_{A_i}}{\partial w_i} = \sum_{j\in N_i}\frac{1}{2\left(g_j - g_i\right)}n_{ij}\int_{\Delta_{ij}} f(x)\ dx \tag{16}$$

Thus, the first term of the (13) is calculated.

Similarly, the term $\frac{\partial f_{A_j}}{\partial w_i}$ in (13) is calculated as follows:

$$\frac{\partial f_{A_j}}{\partial w_i} = \int_{\Delta_{ij}}\left(\frac{\partial x}{\partial w_i}\cdot n_{ji}(x)\right)f(x)\ dx \tag{17}$$

Substituting (15) into (17) and considering $n_{ij}$=-$n_{ji}$:

$$\frac{\partial f_{A_j}}{\partial w_i} = -\frac{1}{2\left(g_j - g_i\right)}n_{ij}\int_{\Delta_{ij}} f(x)\ dx \tag{18}$$

where $n_{ij} = (g_j - g_i)/||g_j - g_i||$. The partial derivative of $H$ with regard to $w_i$ is produced by substituting (16) and (18) into (13). ■

These theorems prove that a fair partition always exists and can be achieved through the proposed distributed update law. This provides a formal guarantee that the restoration workload can be equitably distributed across regions, promoting fairness in customer outage experiences.

*B. RRC Mathematical Formulation*

*1) Objective Function*

This section outlines the optimization framework for addressing the Distribution System Repair and Restoration Problem (DSRRP). Building on state-of-the-art formulations in literature, the restoration model adopts the standard DSRRP, as presented in studies [19], [20], [21], [22]. The objectives are twofold: (i) to maximize the restored load in distribution systems, and (ii) to minimize the total travel distance between damaged components and depots. The proposed objective function is expressed as (19):

$$max\left(\alpha_1\sum_{\forall t}\sum_{\forall l}\rho_{l,t}p_{l,t}^D - \alpha_2\sum_{\forall t}\sum_{\forall m}d(dep\ ,m)\right) \tag{19}$$

Each term in (19) is assigned a weight factor, $\alpha$. $d(dep\ ,m)$ expresses the distance between the depot and damaged components. $p_{l,t}^D$ is active demand at bus $l$ at time $t$. The binary variable $\rho_{l,t}$ indicates whether the lost load has been restored, with $\rho_{l,t}$=1 indicating that the load is picked up.

It is worth noting that the proposed framework is independent of the specific restoration problem formulation. The fairness-driven partitioning and receding-horizon routing strategies can be readily extended to a broad range of power system restoration models with different objectives and operational constraints. This demonstrates the generality, flexibility, and adaptability of the proposed approach across diverse restoration scenarios.

*2) Distribution System Modeling*

The operational constraints of the distribution network are captured by (20) through (28). The restoration problem is formulated using the well-established linearized DistFlow model [20], [23], which has been widely applied in distribution system restoration studies. The detailed formulation is presented in (20)-(28):

$$0 \le p_{l,t}^{DG} \le p_l^{DG_max}, \forall l,t \tag{20}$$

$$0 \le q_{l,t}^{DG} \le q_l^{DG_max}, \forall l,t \tag{21}$$

$$-u_{k,t}^L P_k^{L_max} \le P_{k,t}^L \le u_{k,t}^L P_k^{L_max}, \forall k,t \tag{22}$$

$$-u_{k,t}^L Q_k^{L_max} \le Q_{k,t}^L \le u_{k,t}^L Q_k^{L_max}, \forall k,t \tag{23}$$

$$\sum_{\forall k\in K(.,l)} P_{k,t}^L + p_{l,t}^{DG} = \sum_{\forall k\in K(l,.)} P_{k,t}^L + \rho_{l,t}p_{l,t}^D, \forall l,t \tag{24}$$

$$\sum_{\forall k\in K(.,l)} Q_{k,t}^L + q_{l,t}^{DG} = \sum_{\forall k\in K(l,.)} Q_{k,t}^L + \rho_{l,t}q_{l,t}^D, \forall l,t \tag{25}$$

$$V_{j,t} - V_{i,t} + \frac{R_k P_{k,t}^L + X_k Q_{k,t}^L}{V_1} \le \left(1 - u_{k,t}^L\right)M\ , \forall k,t \tag{26}$$

$$-\left(1 - u_{k,t}^L\right)M \le V_{j,t} - V_{i,t} + \frac{R_k P_{k,t}^L + X_k Q_{k,t}^L}{V_1}\ , \forall k,t \tag{27}$$

$$1 - \varepsilon \le V_{l,t} \le 1 + \varepsilon, \forall l,t \tag{28}$$

Constraints (20) and (21) delineate the limits on active and reactive power outputs of distributed generators (DGs), respectively. By applying a binary variable $u_{k,t}^L$ to the line flow limitations, constraints (22) and (23) guarantee that power flow through a damaged line is zero. To enable network reconfiguration, line switches can be turned ON or OFF using $u^L_{k,t}$. Constraints (24) and (25) deal with the active and reactive power balance, respectively. Voltage drop relationships are captured by constraints (26)–(27), where a big $M$ formulation is employed to decouple voltage variables when a line is disconnected. Constraint (28) enforces voltage magnitude at all buses remains within the allowable operating range.

*3) Radiality Constraints*

Constraints (29)–(31) collectively enforce a radial (tree-structured) topology for the distribution network during the restoration process. The formulation is based on a spanning-tree representation, which guarantees that the energized network remains connected and free of loops, regardless of the direction of power flow. Constraint (29) links the line-status variable to a unique parent–child relationship, ensuring that each energized line contributes exactly one directed edge. Constraint (30) designates substations as root nodes by prohibiting them from having parent buses, while constraint (31) limits each non-substation bus to at most one parent.

$$\beta_{i,j,t} + \beta_{j,i,t} = u^L_{k,t}, \quad \forall k,t \tag{29}$$

$$\beta_{i,j,t} = 0, \quad \forall j \in \beta_s, \forall i,t \tag{30}$$

$$\sum_i \beta_{i,j,t} \le 1, \quad \forall j,t \tag{31}$$

*4) Energy Storage System (ESS) Modeling*

Energy Storage Systems (ESS) can significantly enhance the restoration process by providing fast-response power support and improving system flexibility during outage conditions. In the proposed framework, ESS units are connected at distribution buses and can operate in either charging or discharging mode at each time step. The charging and discharging limits of ESS are modeled as [24], [25]:

$$0 \le P^{\mathrm{ESS,in}}_{l,t} \le P^{\mathrm{ESS,in,max}}_{l} \tag{32}$$

$$0 \le P^{\mathrm{ESS,out}}_{l,t} \le P^{\mathrm{ESS,out,max}}_{l} \tag{33}$$

$$0 \le E^{\mathrm{ESS}}_{l,t} \le E^{\mathrm{ESS,max}}_{l} \tag{34}$$

To prevent simultaneous charging and discharging, the following constraint is imposed:

$$P^{\mathrm{ESS,in}}_{l,t} \cdot P^{\mathrm{ESS,out}}_{l,t} = 0 \tag{35}$$

The energy evolution of the ESS is described as:

$$E^{\mathrm{ESS}}_{l,t} - E^{\mathrm{ESS}}_{l,t-1} = \left(\eta^{\mathrm{in}} P^{\mathrm{ESS,in}}_{l,t-1} - \eta^{\mathrm{out}} P^{\mathrm{ESS,out}}_{l,t-1}\right)\Delta t \tag{36}$$

where:

- $P^{\mathrm{ESS,in}}_{l,t}$, $P^{\mathrm{ESS,out}}_{l,t}$: charging and discharging power at bus $l$
- $E^{\mathrm{ESS}}_{l,t}$: stored energy
- $P^{\mathrm{ESS,in,max}}_{l}$, $P^{\mathrm{ESS,out,max}}_{l}$: power limits
- $E^{\mathrm{ESS,max}}_{l}$: energy capacity
- $\eta^{\mathrm{in}}$, $\eta^{\mathrm{out}}$: efficiencies

Constraint (35) is convexified and transformed into an MILP-compatible form through the introduction of the binary variable $u^{\mathrm{ESS}}_{l,t}$ and a big-*M formulation*:

$$-u^{\mathrm{ESS}}_{l,t} M \le P^{\mathrm{ESS,in}}_{l,t} \le u^{\mathrm{ESS}}_{l,t} M \tag{37}$$

$$-(1-u^{\mathrm{ESS}}_{l,t}) M \le P^{\mathrm{ESS,out}}_{l,t} \le (1-u^{\mathrm{ESS}}_{l,t}) M \tag{38}$$

where $u^{\mathrm{ESS}}_{l,t} \in \{0,1\}$ ensures that ESS operates in only one mode at each time step.

*5) Modeling of Routing Repair Crews*

Routing Repair Crews (RRCs) play a critical role in the load restoration of distribution networks. For $F$ damaged components, $N = \{0, 1, \dots, j, dp\}$, where 0 and $dp$ are the starting and returning points respectively. RRCs leave the starting point 0 to fix damaged components, then come back to the last point $dp$. Fig.4 shows the routes taken by two RRCs, where RRC1 follows Route1 = {0, *a, b, c*, *dp*}, RRC2 follows Route2 = {*0, d, e, f, g, dp*}, and RRC3 follows Route3 = {0, *h, i, j*, *dp*}. Binary variable $x_{m,n}$ equals one if a crew travels the route $m$ to $n$, otherwise it is zero. The detailed formulation is provided as follows:

**Algorithm 1: Receding Horizon (RH) Policy**

1: **Input**: $\theta \in (0,1)$
2: **IF** *F=0*
3: Let P* be the position that minimizes the total distances to failures serviced in the past ( the initial condition for P* is a random point in A); if $p \neq P^*$, proceed to $P^*$; if not, stop.
4: **else**
5: Repeat
6: Calculate the TSP tour across all of *F*'s failures.
7: Randomly select a $\theta$-fragment of the length of such a TSP tour in a uniform manner, regardless of past data.
8: Beginning from the endpoint nearest to the current position, service the chosen piece.
9: Until *F=0*
10: **End**
11: **Repeat**

$$\sum_{\forall n \in N \setminus \{m\}} x_{m,n} - \sum_{\forall n \in N \setminus \{m\}} x_{n,m} = 0, \quad m,n \in N \setminus \{0, dp\} \tag{39}$$

$$\sum_{\forall m \in N \setminus \{0\}} x_{0,m} = 1, \tag{40}$$

$$\sum_{\forall m \in N \setminus \{dp\}} x_{m,dp} = 1, \quad \forall \sigma \tag{41}$$

$$y_m = \sum_{\forall n \in N \setminus \{0,m\}} x_{m,n}, \quad \forall m \in N \setminus \{dp\} \tag{42}$$

This subsection outlines constraints (40) through (43), representing the routing problem [20]. According to constraint (40), a crew arriving at a damaged component only leaves when the repair is finished. Constraint (41) mandates that crew must leave the depot and constraint (42) demands that crew must return to the depot. Constraint (43) establishes a connection between the binary variables $y_m$ and $x_{m,n}$,where $y_m$ is a binary variable indicating whether a crew visited damaged component $m$. Specifically, if a crew visits a damaged component $m$, then $y_m$ is set to 1.

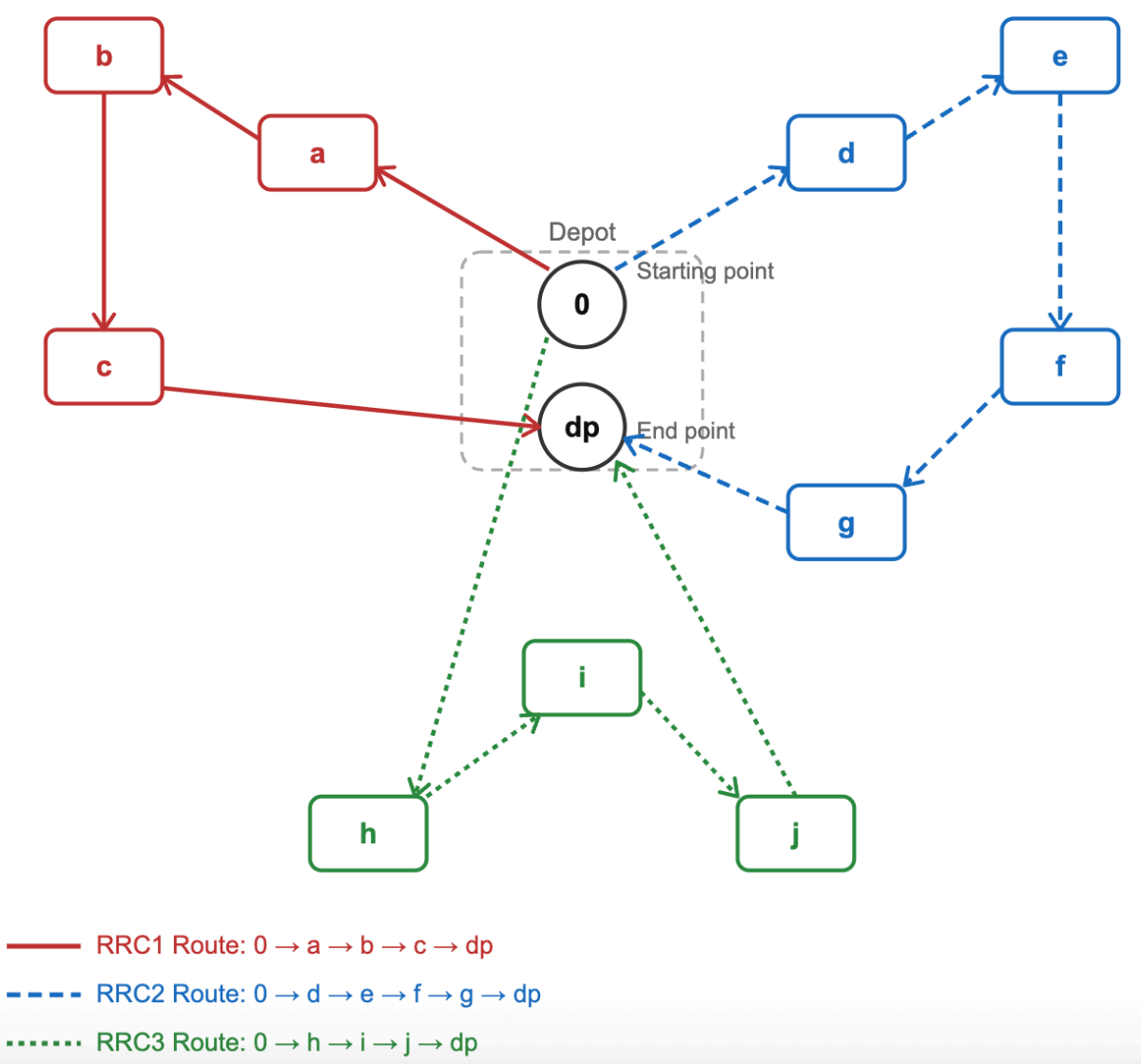

**Fig. 4.** Description of the RRCs

*6) Repair of Damaged Components*

In order to link the Routing Repair Crew's (RRC) arrival times with the repair status of the damaged components, the following arrival constraints are derived. The transition of the distribution line's state from a damaged condition ($u^L_{k,t}$ =0) to a healthy state ( $u^L_{k,t}$= 1) is indicated by the binary variable $u^L_{k,t}$. The following model represents the necessary constraints:

$$AT_m + r_m + tr_{m,n} - AT_n \leq (1 - x_{m,n})M, \quad \forall m \in N \setminus \{dp\}, n \in N \tag{43}$$

$$\sum_{\forall t} RF_{m,t} = 1, \quad \forall, m \in N \tag{44}$$

$$\sum_{\forall t} tRF_{m,t} \geq \sum_{\forall m \in N} (AT_{m,c} + r_m y_m), \quad \forall m \in N \tag{45}$$

$$z_{m,t} \leq \sum_{\tau=1}^{t-1} RF_{m,t}, \quad \forall m \in N, t \tag{46}$$

$$u^L_{k,t} \leq z_{m,t}, \quad \forall m \in N, t \tag{47}$$

The arrival time of the crew is represented by (44). After arriving at the location of a damaged component at time $AT_m$ , the crew spends $r_m$ units of time repairing the damaged component. This is followed by $tr_{m,n}$ units of time travelling to the next damaged component $n$. The arrival time and the necessary repair time define how long it will take to fix a damaged component. For instance, if a crew arrives at the damaged component $m = 1$ at time $AT_1 = 1$, and requires $r_{m1} =$ 3.5 units of time for repair, the component is repaired at $AT_1 + r_{m1} = 4.5$. If the crew does not follow route $x_{m,n}$, the big $M$ approach is used to decouple the arrival timings at components $m$ and $n$. Once a damaged component $m$ is repaired at time $t$, the binary variable $RF_{m,t}$ is set to 1. The restored component is made available in all ensuing time periods, according to constraint (46). Constraints (47) and (48) establish a connection between the RRC and the distribution network operation constraints. If damage has not been restored, $u^L_{k,t}$ is zero.

## IV. Solution Method

In this section, the Receding Horizon (RH) policy is introduced for single-routing repair crews and then extended to solve a multiple-routing repair crew version.

### A. Single-Crew Receding Horizon Policy

The Receding Horizon (RH) policy is described in Algorithm 1.

In this context, $F$ denotes the set of failures awaiting service, and $P$ represents the current position of the repair crew. Furthermore, given a tour of $T$, a $\theta$-fragment of $T$ is defined as a subtour of $T$. If $F$ is not empty, the repair team chooses a random fragment of the TSP tour through the failures in $F$. Keep in mind that the horizon is flexible and can be altered based on how much it will cost to remove any unresolved failures.

### B. Adaptive and Distributed Policies for the Multi-RRC

The prior single-routing repair crews are extended to the multi-routing repair crews in this section.

In the following, the definitions of unbiased and biased policies are introduced.

The locations of randomly selected failures are represented by $X$ in these definitions, and their wait times are indicated by $WT$.

***Definition 4***: Let $\pi$ denote a routing policy that governs how repair crews are assigned to service failures across area $A$. If every pair of sets $A_1, A_2 \subseteq A$ has the property $E[WT \mid X \in A_1] = E[WT \mid X \in A_2]$, then a policy $\pi$ is said to be spatially unbiased. On the other hand, if there are sets $A_1, A_2 \subseteq A$, such that $E[WT \mid X \in A_1] > E[WT \mid X \in A_2]$, then a policy $\pi$ is called spatially biassed.

***Definition 5***: for two measurable functions $f_j : A \rightarrow R^2{}_{>0}$ where $j \in \{1,2\}$, an $nc$-partition $\{A_i\}_{i=1}^{nc}$ is simultaneously fair with respect to $f_1$ and $f_2$ if $\int_{A_i} f_j(x)\,dx = \frac{1}{nc}\int_A f_j(x)\,dx$ for every $i \in I_{nc}$ and $j \in \{1,2\}$.

For any two measurable functions $f_j : A \rightarrow R^2{}_{>0}$ with $j \in \{1,2\}$, Theorem 12 in [26] establishes the existence of a $nc$-partition of $A$ that is simultaneously fair with regard to $f_1$ and $f_2$, with each subset $A_i$ being convex.

In the following, ***Theorem 4*** shows that there exists an optimal unbiased policy for the multi-RRC.

***Definition 6***. A π*-partitioning policy refers to a policy in which the environment is partitioned into $nc$ disjoint regions, and each crew applies the optimal single-crew policy $\pi^*$ independently within its assigned partition.

***Theorem 4***: Given an optimal unbiased policy $\pi^*$ for a single-crew, a π*-partitioning policy is an optimal unbiased policy for the multi-RRC setting if the environment $A$ is divided to $nc$ partitions that are simultaneously fair with respect to $f$ and $f^{1/2}$.

***Proof*** **:** The proof of this theorem consists of two parts: first, demonstrating that if there exists an optimal policy $\pi^*$ for a single crew, then it also serves as the optimal policy for the multi-RRC scenario, which can be found in [27] ; and second, establishing that this optimal policy is unbiased, as follows.

It can be demonstrated that such a multi-RRC policy is indeed unbiased. Under steady-state conditions—where the statistical behavior of the system has stabilized over time, let $X$ denote the location of a randomly chosen failures, $T$ represent their system time, and $A_s$ be an arbitrary service region subset in which they occur. In the multi-RRC policy under consideration, the entire region $A$ is partitioned into $nc$-disjoint subregions $\{A_i\}_{i=1}^{nc}$, with each subregion assigned to a dedicated crew that follows a single-crew routing policy. To evaluate whether the multi-RRC policy is spatially unbiased, the conditional expectation $E[T \mid X \in A_s]$ is introduced, which represents the average system time experienced due to failures located in an arbitrary region $A_s$. Using the law of total expectation, this conditional expectation is decomposed over the intersection between $A_s$ and each service subregion $A_{nc}$:

$$E[T \mid X \in A_s] = E[E[T \mid X \in A_s \cap A_i] \mid X \in A_s] \quad (48)$$

Now, assume that each vehicle executes a single-vehicle policy $\pi^*$ that is known to be spatially unbiased. That is, the expected system time is uniform across all locations within its assigned subregion. This implies that the inner conditional expectation is constant throughout $A_i$, regardless of where within $A_i$ the failure occurs. Therefore, we can write:

$$E[T \mid X \in A_s \cap A_i] = E[T \mid X \in A_i] = T_{\pi^*} \quad (49)$$

Since $T_{\pi^*}$ is a constant independent of $A_i$, the outer expectation also yields the same value:

$$[T \mid X \in A_s] = T_{\pi^*} \quad (50)$$

Finally, because the subset $A_s \subseteq A$ was selected arbitrarily, the result holds for any region within the service area. Therefore, the expected system's time is independent of spatial location, confirming that the multi-RRC policy is spatially unbiased. ■

The RH policy and ***Theorem 4*** propose the distributed multi-RRC version of the RH policy outlined in ***Algorithm 2***: 1) the crews, in a distributed manner, divide the environment A into $nc$ openly disjoint subregions $A_i, i \in \{1,2,\dots,nc\}$, whose union constitutes $A$ with respect to $f$ and $f^{1/2}$ 2) the crews assign themselves to the subregions, ensuring a one-to-one correspondence between crews and subregions; and 3) each crew executes the single-RRC version of the RH policy to address the failures within its respective subregion.

| **Algorithm 2: Multivehicle RH Policy** | |
|---|---|
| 1: | The distributed partitioning method is executed by the crews using measure *f* as an input. |
| 2: | Assign one crew per subregion |
| 3: | Within its own subregion, each crew carries out the single-vehicle RH policy simultaneously. |

## V. Simulation and Results

This section investigates the performance of the RH policy under heavy failure load conditions through simulation studies. The system used for these experiments consists of three connected modified IEEE 69-bus distribution networks, as described in [28]. The underlying restoration formulation is consistent with state-of-the-art formulations in [19], [20], [21], ensuring comparability with prior work. A uniform spatial density is assumed for the failure distribution in the simulations. In all simulations five repair crews are deployed, and the performance is assessed across multiple criteria, including partition balance, wait time, restored power, and fairness.

### *A. Evaluation of Partitioning Algorithm*

This scenario focuses on evaluating the effectiveness of the proposed human-aware fair partitioning algorithm in contrast with a traditional nearest-depot-based partitioning algorithm [20]. The goal is to evaluate how each strategy partitions the service area and distributes workload among repair crews.

Fig. 5 shows the initial partitioning result based on the nearest-depot policy, in which the region is partitioned solely according to the Euclidean distance from depots—commonly referred to as a distance-based approach. Since this method does not account for the stochastic nature of outage distributions, it can lead to uneven service assignments. In contrast, Fig. 6 presents the fair partitions produced by the algorithm proposed in this paper. This method adaptively defines partitions based on the existing and anticipated failures, creating partitions that equalize the expected restoration workload across all crews. As seen in Fig. 5 and 6, the fair strategy produces meaningful adjustments in regional boundaries. An illustrative example of this adjustment is observed at the boundary between areas$A_3$ and $A_4$, where it is realigned to achieve a more balanced distribution of failures. Unlike the nearest-depot approach, which prioritizes minimizing travel distance, the proposed method dynamically adjusts service regions to uphold the principle of human-aware fairness, ensuring that no region or community is disproportionately affected by restoration wait times.

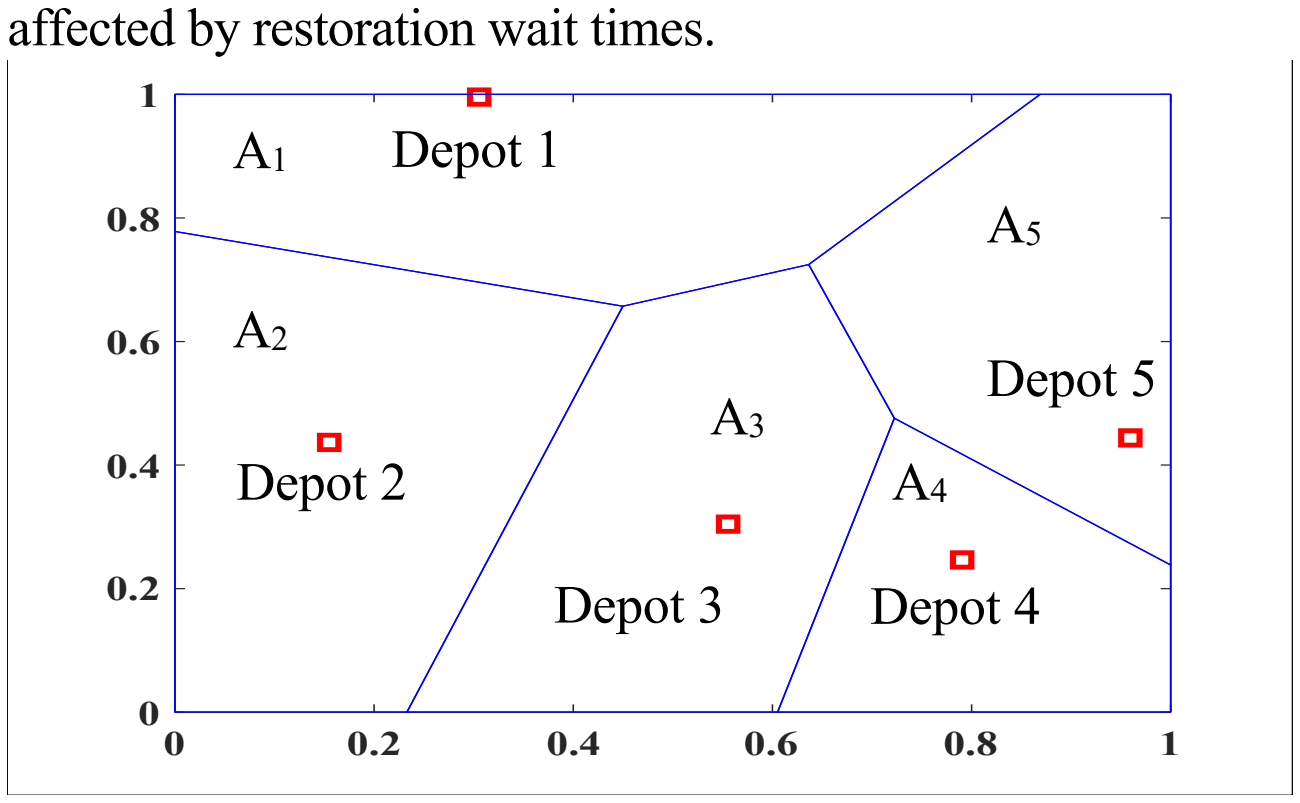

**Fig. 5.** The partitions of the case study based on the nearest-depot method.

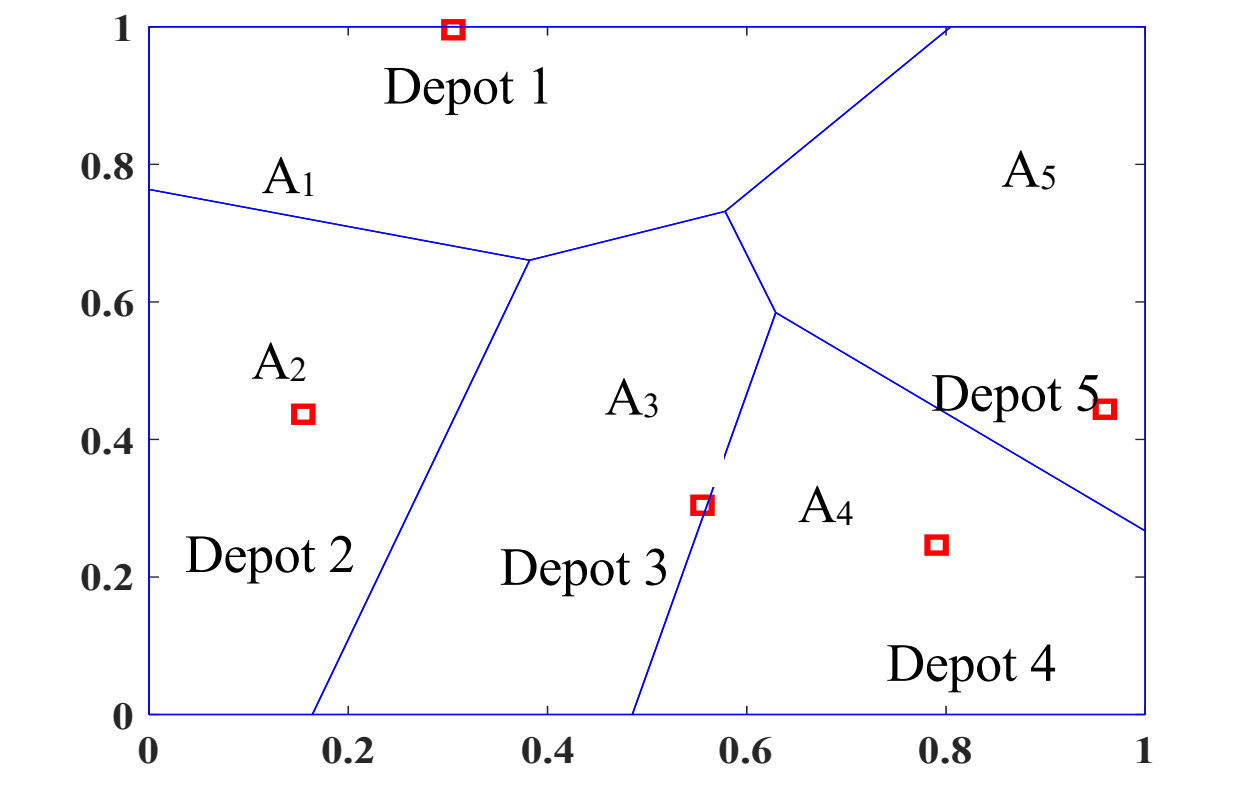

**Fig. 6.** The fair partitions of the case study using the proposed algorithm

### *B. Performance Comparison*

This section presents a comparative analysis between the proposed equity-based partitioning strategy and the conventional distance-based partitioning strategy. The objective is to evaluate how each approach performs in terms of wait time, number of failures per partition, and overall system restoration outcomes under a failure load factor of *FL = 0.99*. TABLE I summarizes the results. The equity-based partitioning strategy achieves a lower maximum wait time of 4.429 compared to the one of distance-based clustering method (i.e. 4.545). Although the minimum wait time is slightly higher in the equity-based method, 4.191 versus 3.99 the standard deviation of wait times is substantially smaller than the one in the distance-based method (i.e. 2.87 vs. 6.029). This reduction in variability indicates a more balanced workload among partitions and supports the effectiveness of the proposed method in achieving human-aware fairness.

TABLE I THE WAIT TIME RESULTS FOR LOAD FACTOR 0.99

| Wait time (h) | Distance-based method | Equity-based method |
|---|---|---|
| **Maximum** | 4.545 | 4.429 |
| **Minimum** | 3.99 | 4.191 |
| **Standard Deviation** | 6.029 | 2.87 |

TABLE II THE NUMBER OF FAILURES RESULTS FOR LOAD FACTOR 0.99

| No. of failures | Distance-based method | Equity-based method |
|---|---|---|
| **Maximum** | 116 | 95 |
| **Minimum** | 58 | 86 |
| **Standard Deviation** | 45.8 | 6.9 |

TABLE II presents the distribution of the number of failures occur across partitions for both the distance-based and equity-based methods under FL = 0.99. The distance-based method results in a highly imbalanced distribution, with a maximum of 116 and a minimum of 58 failures, yielding a standard deviation of 45.7329. This large disparity indicates an over-concentration of failures in certain partitions, resulting in uneven crew workloads and prolonged restoration wait times for affected customers. In contrast, the equity-based method achieves a more uniform distribution of failures, with a maximum of 95, minimum of 87, and a standard deviation of only 6.892. The low standard deviation reflects the algorithm's ability to dynamically partition the network based on probabilistic failure density rather than geographic proximity alone. The result is a more balanced and fair workload allocation, reducing the risk of overburdened crews, and improving overall customer experience.

Fig. 7 presents the cumulative restored power for each of the five partitions under two restoration strategies: (a) the proposed equity-based method and (b) the distance-based clustering approach. In subplot (a), each partition shows a consistent and incremental increase in restored power over time, with minimal divergence between the regions. This close alignment indicates that the proposed algorithm successfully maintains a balanced distribution of restoration efforts, ensuring that no area receives disproportionately delayed service. In contrast, subplot (b) reveals more pronounced disparities among the partitions when the distance-based method is used, particularly in the later stages of restoration. These differences highlight the effectiveness of the equity-based strategy's capability to integrate spatial fairness and load balance, supporting the goal of human-aware fairness throughout the restoration process. Fig. 8 illustrates the cumulative unserved load in percentage over a 30-day period, comparing the equity-based and distance-based restoration scenarios.

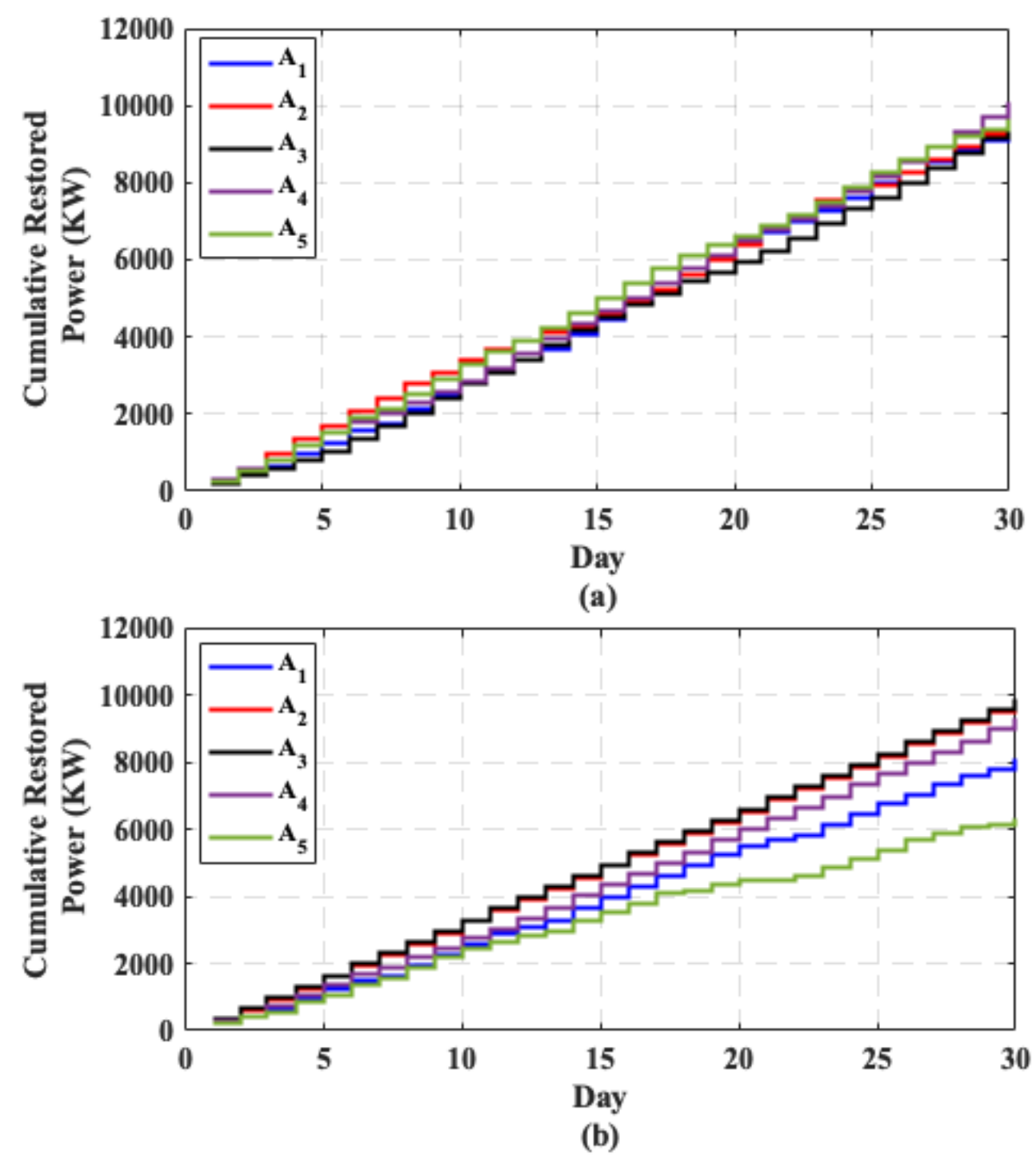

**Fig. 7.** Cumulative restored power for each partition: (a) proposed equity-based partitioning method; (b) Distance-based partitioning method.

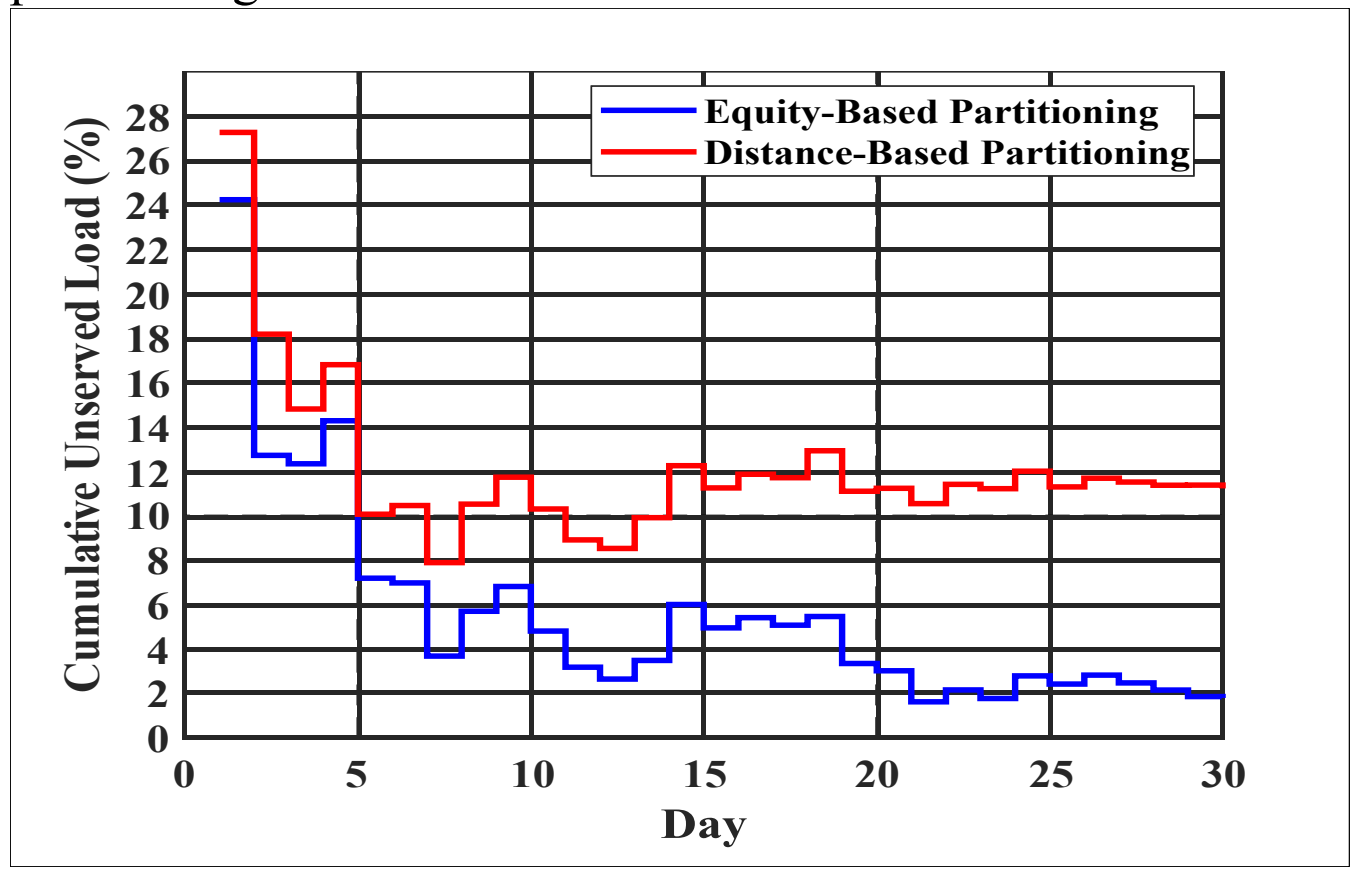

**Fig. 8.** Cumulative unserved load in percentage due to power outage incidents.

The equity-based approach results in decreasing cumulative unserved load, reaching approximately 2% by the end of the simulation period. In contrast, the distance-based method ends at around 12%, indicating significantly less effective load restoration. These results highlighting the effectiveness of the equity-based strategy in accelerating service restoration and achieving more load restoration in a given time period. Together, these findings support the superiority of the equity-based strategy in achieving human-aware fairness and more effective power restoration.

### *C. Impact of Failure Load (FL) Levels on System Performance*

To further evaluate system robustness, the proposed framework is tested under varying failure load factors: *FL = 0.9, 0.95,* and *0.99,* with the parameter $\theta$ set to *0.60***.** As shown in Fig.

9, an increase in the failure load factor reduces the standard deviation in both wait time and the number of failures in different partitions.

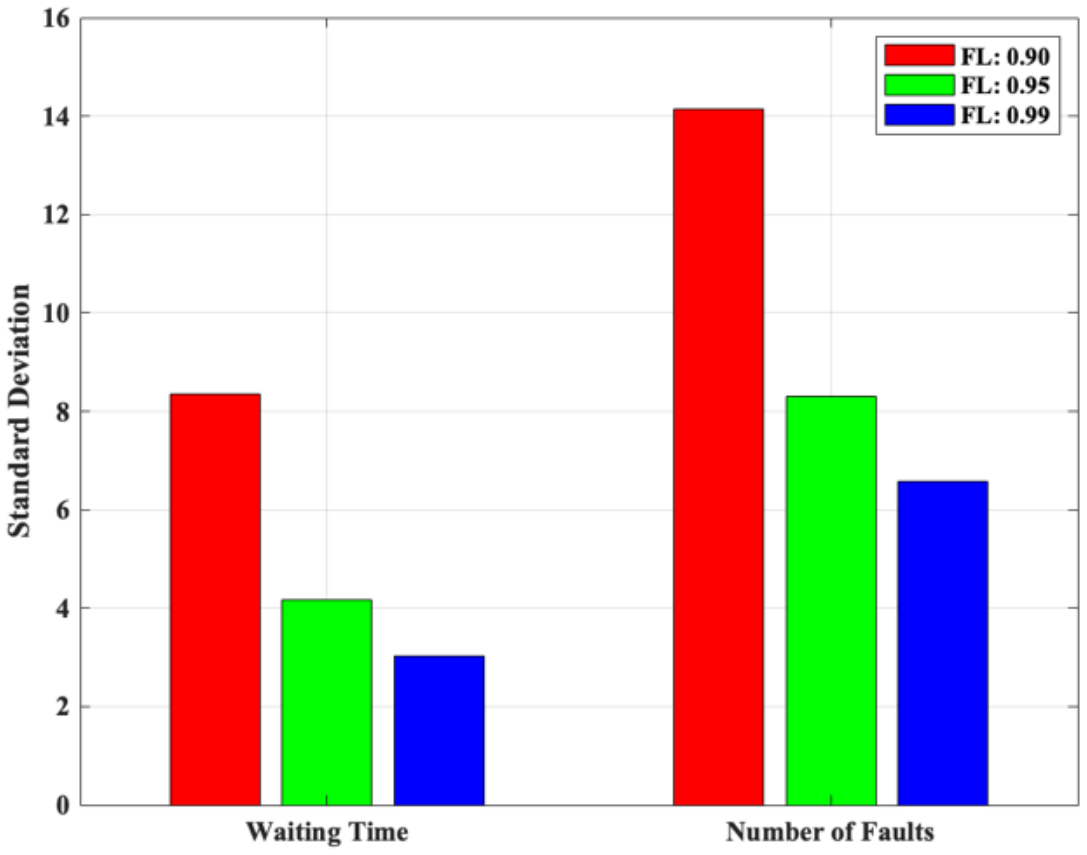

**Fig. 9.** Standard deviation of wait time and number of failures under different failure load factors.

## VI. Conclusion

In this paper, a two-step optimization approach for the repair and restoration in distribution networks was proposed. The first step presented a spatially distributed and adaptive policy designed to create fair restoration partitions, aiming to minimize restoration wait times while ensuring a fair outage experience for customers. This approach overcomes the challenge of non-deterministic failures by incorporating the probability of future failures into the partitioning algorithm. In the second step, an adaptive and distributed algorithm was employed to address failures stochasticity with respect to arrival times, locations, and service requirements. Simulations conducted across various scenarios demonstrated that the proposed method effectively addresses repair and restoration problems under uncertain conditions, ensuring that customers experience approximately equal wait times for service restoration.